\documentclass[aps,prd,twocolumn,showpacs]{revtex4}
\usepackage{graphicx}
\usepackage{dcolumn}
\begin{document}
\hyphenation{Drever Vyatchanin Braginsky Thorne Kimble Matsko}

\title{Sagnac Interferometer as a Speed-Meter-Type, Quantum-Nondemolition Gravitational-Wave
Detector}

\author{Yanbei Chen}

\affiliation{Theoretical Astrophysics, California Institute of Technology, Pasadena,
CA 91125}

\begin{abstract}
According to quantum measurement theory, ``{\it speed meters}'' --- devices that measure the
momentum, or speed, of free test masses --- are
immune to the standard quantum limit (SQL). It is shown that a Sagnac-interferometer
gravitational-wave detector is a speed meter and therefore in principle it
can beat the
SQL by large amounts over a wide band of frequencies.  
It is shown, further, that, when one ignores optical losses,
a signal-recycled Sagnac interferometer with
Fabry-Perot arm cavities
has precisely the same performance, for the same circulating light power,
as the Michelson speed-meter interferometer recently invented and
studied by P. Purdue and the author.  The influence of optical losses
is not studied, but it is plausible that they be fairly unimportant for the
Sagnac, as for other speed meters.  
With squeezed vacuum (squeeze factor $e^{-2R} = 0.1$) injected into its dark
port, the recycled Sagnac
can beat the SQL by a factor $ \sqrt{10} \simeq 3$ over the frequency band
$10 {\rm Hz} \alt f \alt 150 {\rm Hz}$ using
the same circulating power $I_c\sim 820$ kW as is used by the (quantum limited) 
second-generation Advanced LIGO
interferometers --- if other noise sources are made sufficiently small.  It is
concluded that the Sagnac optical configuration, with signal recycling
and squeezed-vacuum injection, is
an attractive candidate for third-generation interferometric
gravitational-wave detectors (LIGO-III and EURO).
\end{abstract}

\maketitle

\section{Introduction}

After decades of planning and development, an array of large-scale
laser interferometric gravitational-wave detectors (interferometers
for short), consisting of the Laser Interferometer Gravitational-wave
Observatory (LIGO), Virgo, GEO and TAMA \cite{GWarray}, is gradually
becoming operative, targeted at gravitational waves in the high-frequency
band ($ 10$--$ 10^{3}\, \mathrm{Hz}$). Michelson-type laser
interferometry is used in these detectors to monitor 
gravitational-wave-induced changes in
the separations of mirror-endowed test masses.  More specifically,  
a laser beam is split in two by a 50/50
beamsplitter, and the two beams are sent into the two arms (which may contain
Fabry-Perot cavities) and then brought back together and interfered,
yielding a signal that senses the difference
of the two arm lengths. Although it is plausible that gravitational
waves will be detected, for the first time in history, by these initial
interferometers, a significant upgrade of them must probably be 
made before a rich program of  
observational gravitational-wave astrophysics
can be carried out \cite{matureLIGO}. In the planned upgrade of the
LIGO interferometers (\emph{Advanced LIGO}, tentatively
scheduled to begin operations 
in $ 2008$ \cite{whitpaper}), the Michelson topology will still be used,
as also is probably the case for Advanced LIGO's international counterparts,
for example the Japanese LCGT \cite{LCGT}. 

An alternative to the Michelson topology, the Sagnac topology, originally
invented in 1913 \cite{SagnacOrig}
for rotation sensing, can also be used for gravitational-wave
detection \cite{Drever,Sagnac}. In a Sagnac interferometer, as in a Michelson,
a laser beam
is split in two, but each of the two beams travels successively through
both arms, though in the opposite order (in opposite directions).
When the
two beams are finally recombined, a signal sensitive to the
\emph{time-dependent part} of the arm-length difference is obtained.

Until now, there has been little motivation to 
switch from the more mature Michelson topology to the Sagnac topology,
because: (i) the technical advantages provided by the Sagnac topology
have not been able to overcome its disadvantages \cite{Sagnac,SGM98,PGM98},
and (ii) the \emph{shot-noise limited} sensitivities of ideal Sagnac
interferometers have not exhibited any interesting features \cite{MRSWD97}.
Nevertheless, a sustained research effort is still being made on the
Sagnac topology, aimed at third generation gravitational-wave
detectors (beyond Advanced LIGO). In particular, an all-reflective
optical system suitable for the Sagnac is being developed
\cite{AllReflective}, with the
promise of being able to cope with the very high laser powers 
that may be needed in the third generation. 

In this paper, a theoretical study of the idealized noise performance
of Sagnac-based interferometers at high laser powers is carried out.
It is shown that, by contrast with the previously studied
low-power regime, the (ideal) Sagnac interferometer
might be significantly better at high powers than its ideal Michelson 
counterparts, and thus is an attractive candidate for
third-generation interferometric gravitational-wave
detectors, e.g., LIGO-III and EURO \cite{EURO}.

In advanced gravitational-wave interferometers, the laser power is
increased to lower the shot noise. However, at these higher light powers,
the photons in the arms
exert stronger random forces on the test masses, thereby inducing
stronger \emph{radiation-pressure noise}. At high enough laser powers
(above about 850 kW in Advanced LIGO),
the radiation-pressure noise can become larger than the shot noise
and dominate a significant part of the noise spectrum (usually at
all frequencies below the noise-curve minimum). As was first pointed out 
by Braginsky in the 1960s \cite{VB60s,BK92},
a balance between the two noises gives rise to a Standard Quantum
Limit (SQL).  As was later realized, again by Braginsky \cite{VB60s,BK92},
the SQL can be circumvented by clever designs, which he named Quantum
Non-Demolition (QND) schemes. 

The advanced LIGO interferometers were originally planned to operate 
near or at the
SQL \cite{whitpaper}, but it was later shown by Buonanno and
Chen that they can actually beat the SQL by a moderate 
amount over a modest frequency
band, due to a change
in interferometer dynamics \cite{BC1,BC2,BC3} induced by detuned
signal-recycling \cite{SR,RSE}. 

Generations beyond Advanced LIGO,
however, will have to beat the SQL by significant amounts over a broad
frequency band; i.e., they must be {\em strongly QND}. Currently
existing schemes for strongly QND interferometers with Michelson topology 
include:
(i) The use of two additional kilometer-scale optical filters to perform
frequency-dependent homodyne detection \cite{FDH} at the output
of a conventional Michelson interferometer, as invented and 
analyzed by Kimble,
Levin, Matsko, Thorne and Vyatchanin (KLMTV) \cite{KLMTV}.  (Reference
\cite{KLMTV} can be used as a general starting point for the
quantum-mechanical analysis of QND gravitational-wave interferometers.)
(ii) The speed-meter interferometer, originally invented by Braginsky
and Khalili \cite{SMorig}, developed by Braginsky, Khalili, Gorodetsky
and Thorne \cite{SMBGKT}, and later incorporated into the Michelson
topology by Purdue and Chen \cite{SMP,SMPC}. In its Michelson form, the
speed meter uses at least
one additional kilometer-scale optical cavity to measure the relative momentum
of the free test masses over a broad frequency band. 

The speed meter is motivated, theoretically, by the fact that the momentum 
of a free test mass is a {\em QND observable} \cite{QNDobs} 
i.e., it can be measured continuously
to arbitrary accuracy without being limited by the SQL. 
Practically,
QND schemes based on a Michelson speed meter can exhibit broadband
QND performances using only one additional kilometer-scale cavity, 
by contrast with the two additional cavities needed for 
QND schemes based on a conventional Michelson interferometer
(a {\em position meter}). Michelson--speed-meter-based QND schemes are also
less susceptible to optical losses than those based on
Michelson position meters (Sec.~V of \cite{SMPC}). 

Surprisingly, so far as we are aware nobody has previously noticed that, 
because
the Sagnac interferometer is sensitive only to the \emph{time-dependent
part} of the arm-length difference, it is automatically a speed meter.
Moreover, as we shall see
in this paper, with the help of \emph{signal-recycling} \cite{SR,RSE},
i.e., by putting one additional mirror at its dark output port, a
Sagnac interferometer
can be optimized to have a comparable performance to a Michelson
speed meter, \emph{without the need for any additional kilometer-scale
cavities}. In particular, a signal-recycled Sagnac interferometer
with ring cavities in its arms has exactly the same performance
as the Michelson speed meters of Ref.~\cite{SMPC}, aside from (presumably
minor) differences due to optical losses. 

This paper is organized as follows: in Sec.~II we derive the
input-output relation of signal-recycled Sagnac interferometers, with
either optical delay lines or ring-shaped Fabry-Perot cavities in the
arms, showing that they are indeed measuring the relative speed of test
masses.  In Sec.~III, we evaluate the noise spectral density of ideal Sagnac
interferometers, obtaining comparable performances to the Michelson
speed meters. In Sec.~IV, we discuss some technical issues that deserve further
investigation. Finally, Sec.~V summarizes our conclusions. Appendix A
contains details in the calculations of the input-output relation of a
single interferometer arm, which might contain an optical delay line
or a ring cavity. 

\begin{figure}
\includegraphics[width=0.5\textwidth]{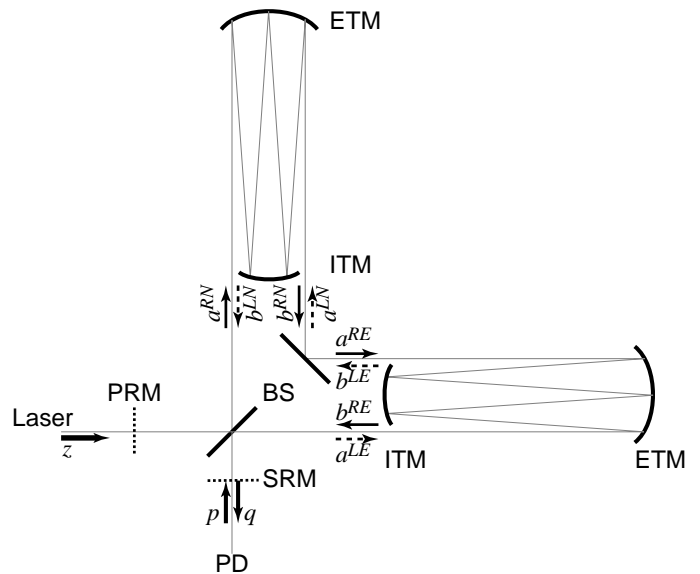}
\caption{Schematic plot of a Sagnac interferometer with optical delay lines
in the arms.
\label{sagnac_DL}}
\end{figure}

\begin{figure}
\includegraphics[width=0.4\textwidth]{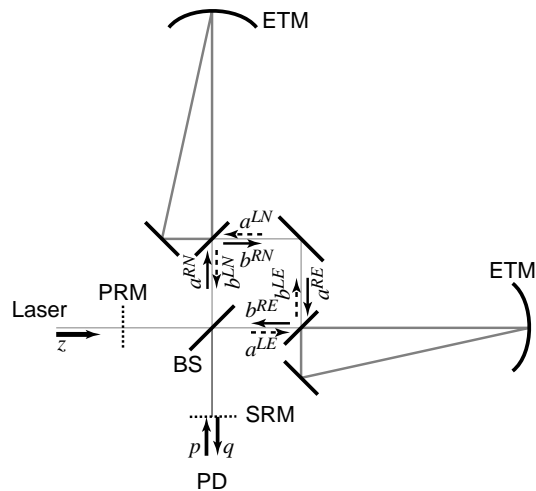}
\caption{Schematic plot of a Sagnac interferometer with ring cavities in the
arms.\label{sagnac_FP}}
\end{figure}

\section{The Sagnac as a speed meter, and its input-output relations}

\subsection{The Sagnac optical configuration}

Two variants of Sagnac interferometers are shown in Figs.~\ref{sagnac_DL}
and \ref{sagnac_FP}, which uses optical delay lines or ring-shaped
Fabry-Perot cavities in the arms. [The idea of using optical delay
  lines and ring cavities are due to  Drever, proposed in similar
  designs. See, e.g., Ref.~\cite{Drever}.]  
In both variants, the carrier light enters the interferometer
from the left port (also called the ``bright port'') of the beamsplitter (BS). 
The light
gets split in two and travels into the two arms in opposite orders; we
denote by R the beam that enters the North (N) arm first and the East
(E) arm second, and by L that which enters E first and N second. 
Suppose the mirrors are all
held fixed at their equilibrium positions, so that the two arms are
exactly symmetric; then the carrier R and L beams,
upon arriving again at the beamsplitter, will combine in such a way
that no lights exits to the port below the beamsplitter 
(the ``dark port''). Similarly, any vacuum fluctuations that enter the
interferometer at the bright port along with the carrier light will 
also be suppressed
in the dark-port output. Only vacuum fluctuations that entered
the interferometer from the dark port can leave the interferometer
through the dark port. As a result, the dark port is decoupled from the
bright port, as in a Michelson interferometer. This fact is crucial
to the suppression of laser noise in the dark-port output. 

\subsection{The Sagnac's Speed-Meter Behavior}

When the end mirrors of the two arms are allowed to move, they 
phase modulate
the carrier light, generating sideband fields. Only antisymmetric,
non-static changes in the arm lengths can contribute to the dark-port
output; this is a result of the cancellation at the beamsplitter, and the
fact that the two beams pass through the two arms in the opposite order. A
more detailed but still rough exploration of this point reveals the
Sagnac's role as a speed-meter interferometer:

Denoting by $ \tau _{\mathrm{arm}}$ the (average)
storage time of light in the arms and by $ x_{\mathrm{N},\, \mathrm{E}}$ the 
time-dependent
displacements of the end mirrors, we have for the phase gained by
the R and L beams after traveling from the bright entry port through the 
two arms to the dark exit port:
\begin{eqnarray*}
\delta \phi _{\mathrm{R}} & \sim & \mathrm{x}_{\mathrm{N}}(t)+x_{\mathrm{E}}(t+\tau _{\mathrm{arm}})\, ,\\
\delta \phi _{\mathrm{L}} & \sim & \mathrm{x}_{\mathrm{E}}(t)+x_{\mathrm{N}}(t+\tau _{\mathrm{arm}})\, .
\end{eqnarray*}
The amplitude of the dark-port output is proportional to the phase
difference of the two beams at the beamsplitter:
\begin{eqnarray} 
\delta \phi _{\mathrm{R}}-\delta \phi _{\mathrm{L}} \nonumber &\sim& \left[
  x_{\mathrm{N}}(t)-x_{\mathrm{N}}(t+\tau
  _{\mathrm{arm}})\right]\nonumber \\
&-&\left[ x_{\mathrm{E}}(t)-x_{\mathrm{E}}(t+\tau
  _{\mathrm{arm}})\right] \, . 
\label{dphiR_dphiL}
\end{eqnarray}
As a consequence, the Sagnac interferometer is \emph{not} sensitive
to any time-independent displacement of the test masses.  By expanding
Eq.\ (\ref{dphiR_dphiL}) in powers of $\tau_{\rm arm}$, we see that,
at frequencies much smaller than $ 1/\tau _{\mathrm{arm}}$,
the \emph{speed} of the test-mass motion is measured, and at higher
frequencies, a mixture of the speed and its time derivatives 
is measured --- as also is the case in other speed meters \cite{SMorig,SMP,SMPC}.

\subsection{Input-output relations without a signal-recycling mirror}

As a foundation for evaluating the performances of Sagnac interferometers
in the high-power regime, we shall now derive their quantum mechanical
{\em input-output relations} --- i.e., we shall derive equations for
the quantum mechanical dark-port output field
$ q$ in terms of the input (vacuum) fields $ p$ at the dark
port and $ z$ at the bright port (which in the end does not appear
in $ q$), and the gravitational-wave field $ h$.  In Figs.~\ref{sagnac_DL}
and \ref{sagnac_FP}, we have denoted by $ a^{\mathrm{RN},\, \mathrm{RE},\, \mathrm{LN},\, \mathrm{LE}}$
the input sideband fields of the R and L beams at the N and E arms,
and by $ b^{\mathrm{RN},\, \mathrm{RE},\, \mathrm{LN},\, \mathrm{LE}}$
the output sideband fields. For the moment, we shall ignore the existence
of the signal-recycling mirror (SRM); and throughout we shall ignore
the power-recycling mirror (PRM) since (as for Michelson topologies) it
merely serves to provide a larger input power at the beamsplitter and has
no other significance for the interferometer's quantum noise.

In this paper, we shall use the Caves-Schumaker two-photon formalism
\cite{Caves_Schumaker} (briefly introduced in Sec.~IIA of KLMTV),
which breaks the time-domain sideband fields, at any given spatial location, into the following form, 
\begin{equation}
E(t)=\sqrt{\frac{4\pi \hbar \omega _{0}}{{\mathcal{A}}c}}\left[
  E_{1}(t)\cos (\omega _{0}t)+E_{2}(t)\sin (\omega _{0}t)\right] \,,\end{equation} 
where $\omega_0$ is the carrier frequency, ${\cal A}$ is the cross
  sectional area of the beam.  Here $ E_{1,2}(t)$ are slowly varying fields called the cosine
(or amplitude) and sine (or phase) quadratures. These quadratures fields
can be thought of as amplitude or phase modulations on a carrier
field of the form $ D\cos (\omega _{0}t)$ . The quadrature fields
can be expanded as

\begin{equation}
E_{1,2}(t)=\int _{0}^{+\infty }\frac{d\Omega }{2\pi }\left(
a_{1,2}e^{-i\, \Omega \, t}+a_{1,2}^{\dagger }e^{+i\, \Omega \,
  t}\right) \, ,
\end{equation}
in terms of the quadrature operators $ a_{1,2}(\Omega )$. A more general
quadrature operator can be constructed from $ a_{1,2}$: 
\begin{equation}
a_{\Phi }=a_{1}\cos \Phi +a_{2}\sin \Phi \, .
\end{equation}

The set of propagation equations common to both of our Sagnac configurations
{[}with either delay lines (DL for short) or ring cavities (FP for
short) inside the arms{]} are: (i) at the beamsplitter,
\begin{equation}
\label{eq:BS}
a^{\mathrm{RN}}=\frac{z+p}{\sqrt{2}}\, ,\quad
a^{\mathrm{LE}}=\frac{z-p}{\sqrt{2}}\, ,\quad
q=\frac{b^{\mathrm{LN}}-b^{\mathrm{RE}}}{\sqrt{2}}\, ; 
\end{equation}
and (ii) when the beams leave one arm and enter the other,
\begin{equation}
\label{eq:connection}
a^{\mathrm{RE}}=b^{\mathrm{RN}}\, ,\quad a^{\mathrm{LN}}=b^{\mathrm{LE}}\, .
\end{equation}
The above equations, (\ref{eq:BS}) and (\ref{eq:connection}),
are for both quadratures. By writing down these equations, we assume the distances between the BS and
ITMs to be small, and also integer multiples of the laser wavelength.

\begin{table*}
\begin{tabular}{|c|c|c|}
\hline 
&
DL&
FP\\
\hline
\hline 
$ \Psi _{\mathrm{arm}}$&
$ \displaystyle \mathcal{B}\Omega L/c$&
$ \displaystyle \arctan \left( \frac{1+\sqrt{R}}{1-\sqrt{R}}\tan \frac{\Omega L}{c}\right) $\\
\hline 
$ \mathcal{K}_{\mathrm{arm}}$&
$ \displaystyle \frac{8I_{\mathrm{c}}\omega _{0}}{m\mathcal{B}\Omega ^{2}c^{2}}\left( \frac{\sin \mathcal{B}\Omega L/c}{\sin \Omega L/c}\right) ^{2}$&
$ \displaystyle \frac{8I_{\mathrm{c}}\omega _{0}}{m\Omega ^{2}c^{2}}\left( \frac{T}{1-2\sqrt{R}\cos (2\Omega L/c)+R}\right) $\\
\hline 
$ \Psi _{\mathrm{sagnac}}$&
$ \displaystyle 2\mathcal{B}\Omega L/c+\pi /2$&
$ \displaystyle 2\arctan \left( \frac{1+\sqrt{R}}{1-\sqrt{R}}\tan \frac{\Omega L}{c}\right) +\pi /2$\\
\hline 
$ \mathcal{K}_{\mathrm{sagnac}}$&
$ \displaystyle \frac{32I_{\mathrm{c}}\omega _{0}}{mLc\left(
  \frac{c}{L\mathcal{B}}\right) ^{3}}\left[ \frac{\sin
    ^{2}(\mathcal{B}\Omega L/c)}{(\mathcal{B}\Omega
    L/c)(\mathcal{B}\sin \Omega L/c)}\right] ^{2}$& 
$ \displaystyle \frac{32I_{\mathrm{c}}\omega _{0}}{m\Omega
  ^{2}c^{2}}\left[ \frac{(1+\sqrt{R})\sqrt{T}\sin (\Omega
    L/c)}{1-2\sqrt{R}\cos (2\Omega L/c)+R}\right] ^{2}$\\  
\hline
\end{tabular}

\caption{Expressions for \protect$ \Psi _{\mathrm{arm}}\protect$, \protect$ \mathcal{K}_{\mathrm{arm}}\protect$,
\protect$ \Psi _{\mathrm{sagnac}}\protect$ and \protect$ \mathcal{K}_{\mathrm{sagnac}}\protect$
in the DL and FP cases. \label{tab:KPsi}}
\end{table*}

The input-output relations for the arms, i.e., the $ b$-$ a$
relations, are evaluated in the Appendix (in an manner analogous to that
of KLMTV for Michelson configurations), 
for the distinct cases of DL and FP. The results can be
put into the following simple form:
\begin{eqnarray}
b_{1}^{IJ} & = & e^{2i\Psi _{\mathrm{arm}}}a_{1}^{IJ}\, ,\label{eq:genericinout1} \\
b_{2}^{IJ} & = & e^{2i\Psi _{\mathrm{arm}}}\left[ 
  a_{2}^{IJ}-\mathcal{K}_{\mathrm{arm}}(a_{1}^{\mathrm{L}J}+a_{1}^{\mathrm{R}J})\right] \nonumber \\
&& +e^{i\Psi
  _{\mathrm{arm}}}\frac{\sqrt{2\mathcal{K}_{\mathrm{arm}}}}{h_{\mathrm{SQL}}}\sqrt{2}\, \tilde{x}_{J}^{\mathrm{GW}}/L\, .\label{eq:genericinout2}          
\end{eqnarray}
Here $I=\mathrm{L},\, \mathrm{R}$ stands for either one of the two beams,
and $J=\mathrm{E},\, \mathrm{N}$ stands for either one of the two
arms. The quantity $\tilde{x}_J^{\rm GW}$ is the gravitational-wave induced
displacement of the $J$th ETM (in frequency domain), $L$ is the arm length, and $m$ is the
mass of the ITM and the ETM. The Standard Quantum Limit is given by 
\begin{equation}
\label{hsql}
h_{\mathrm{SQL}}=\sqrt{\frac{8\hbar }{m\Omega ^{2}L^{2}}}\,.
\end{equation}
 Expressions for $ \Psi _{\mathrm{arm}}$ and $ \mathcal{K}_{\mathrm{arm}}$,
in the cases of DL and FP, are given in the Appendix {[}Eqs.~(\ref{eq:PsiDLapp}),
(\ref{eqKDLapp}), (\ref{eq:PsiKFPapp1}) and (\ref{eq:PsiKFPapp2}){]}
and summarized in Table.~\ref{tab:KPsi}. 
Combining Eqs.~(\ref{eq:BS})--(\ref{eq:genericinout2}),
we obtain $ q_{1,2}$ in terms of the input fields and the 
dimensionless gravitational-wave strain (in frequency domain), $\tilde{h}$ {[}also using $
  \tilde{h}=(\tilde{x}_{\mathrm{E}}^{\mathrm{GW}}-\tilde{x}_{\mathrm{N}}^{\mathrm{GW}})/L${]}:
\begin{eqnarray}
q_{1} & = & e^{2i\Psi _{\mathrm{sagnac}}}p_{1}\, ,\label{eq:inoutsagnacgeneric1} \\
q_{2} & = & e^{2i\Psi
  _{\mathrm{sagnac}}}(p_{2}-\mathcal{K}_{\mathrm{sagnac}}p_{1})
\nonumber \\
&+& e^{i\Psi _{\mathrm{sagnac}}}\frac{\sqrt{2\mathcal{K}_{\mathrm{sagnac}}}}{h_{\mathrm{SQL}}}\tilde{h}\, ;\label{eq:inoutsagnacgeneric2}    
\end{eqnarray}
 with
\begin{eqnarray}
\label{eq:PsiKsagnac1}
\Psi _{\mathrm{sagnac}}&=&2\Psi _{\mathrm{arm}}+\frac{\pi }{2}\,,\\
\label{eq:PsiKsagnac2}
 \mathcal{K}_{\mathrm{sagnac}}&=&4\mathcal{K}_{\mathrm{arm}}\sin ^{2}\Psi _{\mathrm{arm}}\, .
\end{eqnarray}
Expressions for $ \Psi _{\mathrm{sagnac}}$ and $ \mathcal{K}_{\mathrm{sagnac}}$
in the DL and FP cases can be obtained by inserting Eqs.~(\ref{eq:PsiDLapp}),
(\ref{eqKDLapp}), (\ref{eq:PsiKFPapp1}) and (\ref{eq:PsiKFPapp2})
into Eqs.~(\ref{eq:PsiKsagnac1}) and(\ref{eq:PsiKsagnac2}), with results summarized again in
Table.~\ref{tab:KPsi}. Indeed, as mentioned at the
beginning of this section, the bright-port input field $z$ does not
appear in the dark-port output quadratures, $q_{1,2}$. 

The
input-output relations (\ref{eq:inoutsagnacgeneric1})
and (\ref{eq:inoutsagnacgeneric2}) have the same general form as those of
a conventional Michelson interferometer, Eq.~(16) of \cite{KLMTV},
and those of a Michelson speed meter, Eqs.~(27) of \cite{SMP} or Eqs.~(12)
of \cite{SMPC}. In particular, as discussed in the Appendix, the
output phase quadrature $ q_{2}$ is a sum of three terms: the
shot noise (first term), the radiation-pressure noise (second term)
and the gravitational-wave signal (third term), while the output amplitude
quadrature $ q_{1}$ contains only shot noise. 
\subsection{Influence of signal recycling on the input-output relations}

Since the input-output relations of Sagnac interferometers have the
same form as those of a conventional Michelson interferometer, the
quantum noise of signal-recycled Sagnac interferometers can be obtained
easily using the results of Refs.~\cite{BC1,BC2}. For simplicity,
we shall restrict the signal-recycling cavity to be either resonant 
with the carrier frequency (``tuned
SR'') or anti-resonant (``tuned RSE''), leaving the detuned case for
future investigations.  In these
cases, the dynamics of the interferometer are not modified
by the signal recycling, and the
input-output relation has the same form as Eqs.~(\ref{eq:inoutsagnacgeneric1})
and (\ref{eq:inoutsagnacgeneric2}), with $ \mathcal{K}_{\mathrm{sagnac}}$
replaced by (see Sec.~IIIC of Ref.~\cite{BC1})
\begin{equation}
\label{eq:KsagnacSR}
\mathcal{K}_{\mathrm{sagnac}\, \mathrm{SR}}=\frac{\tau ^{2}}{1-2\rho
  \cos 2\Psi _{\mathrm{sagnac}}+\rho
  ^{2}}\mathcal{K}_{\mathrm{sagnac}}\,,
\end{equation}
and $\Psi_{\rm sagnac}$ replaced by a quantity $\Psi_{\rm sagnac\,SR}$
  whose value is not of interest to us. Here $\rho$ and $\tau$ are the (amplitude) reflectivity and
transmissivity of the signal-recycling mirror, with $\rho\in\Re$,
  $\tau>0$ and 
$\rho^2+\tau^2=1$. The quantity $\phi$ is the {\it detuning} of the 
signal-recycling cavity, related to its length by $l$ (which is
  defined  as the distance from the beamsplitter to the
  signal-recycling mirror) by $\phi=[\omega_0
  l/c]_{\rm mod\,2\pi}$. It is also assumed that the length $l$ is
very short, so $\Omega l/c$ can be regarded as $0$.  Expressions for $ \mathcal{K}_{\mathrm{sagnac}\, \mathrm{SR}}$
can be obtained by using results in Table \ref{tab:KPsi}. 

Using the
fact that $ \Omega L/c\ll 1$ (for earth-based interferometers in the
high-frequency band), $ \mathcal{B}\gg 1$ (for the DL
case) and $ T\ll 1$ (for the FP case), we can obtain some approximate
formulas for $ \mathcal{K}_{\mathrm{sagnac}\, \mathrm{SR}}$ (which
also apply to the non-SR case, with $\rho\rightarrow 0$ and
$\tau\rightarrow 1$): in
the DL case
\begin{eqnarray}
\label{eq:KSRapp}
\mathcal{K}_{\mathrm{sagnac}\,
  \mathrm{SR}}^{\mathrm{DL}}=&&\frac{32I_{\mathrm{c}}\omega
  _{0}}{mLc\gamma _{\mathrm{DL}}^{3}}
\left[ \frac{\tau ^{2}}{1+2\rho
    \cos (4\Omega /\gamma _{\mathrm{DL}})+\rho ^{2}}\right] 
\nonumber\\
&&\times\left[
  \frac{\sin (\Omega /\gamma _{\mathrm{DL}})}{\Omega /\gamma
    _{\mathrm{DL}}}\right] ^{4}\, , 
\end{eqnarray}
with
\begin{equation}
\label{eq:gammaDL}
\gamma _{\mathrm{DL}}=\frac{c}{{\mathcal{B}}L}\, ;
\end{equation}
 and in the FP case

\begin{equation}
\label{eq:KFPsagnacSR}
\mathcal{K}_{\mathrm{sagnac}\,
  \mathrm{SR}}^{\mathrm{FP}}=\frac{16I_{\mathrm{c}}\omega
  _{0}}{mLc}\frac{\delta }{(\Omega ^{2}-\Omega
  _{\mathrm{s}}^{2})^{2}+\delta ^{2}\Omega ^{2}}\, ; 
\end{equation}
with
\begin{equation}
\label{eq:delta}
\delta = 2\left( 1+\frac{T}{2}\right) \frac{1-\rho }{1+\rho
}\gamma _{\mathrm{FP}}\, \quad
\Omega _{\mathrm{s}}=\left(
1+\frac{T}{2}\right) \gamma _{\mathrm{FP}}\,,
\end{equation}
where
\begin{equation}
\label{eq:gammaFP}
\gamma_{\mathrm{FP}}=\frac{Tc}{4L}\, . 
\end{equation} 
Interestingly, Eq.~(\ref{eq:KFPsagnacSR}) is identical to Eqs.~(22)
and (23) of Ref.~\cite{SMPC}, with substitutions (this paper $ \rightarrow $
Purdue and Chen) $ I_{c}\rightarrow W_{\mathrm{circ}}$ (circulating
power), $ \Omega _{\mathrm{s}}\rightarrow \Omega $ (sloshing frequency),
$ \delta \rightarrow \delta $ (extraction rate), $ \Omega \rightarrow \omega $
(sideband frequency), and $ \mathcal{K}_{\mathrm{sagnac}\, \mathrm{SR}}^{\mathrm{FP}}\rightarrow \kappa $.
As we shall explain further in the following sections, the coupling constant
$ \mathcal{K}_{\rm sagnac\,SR}$ 
alone (besides $h_{\rm SQL}$, which depends on $m$ and $L$) will
determine the quantum noise of the interferometer. 
This means that a signal-recycled Sagnac interferometer with ring cavities
in its arms is equivalent to the Michelson speed meters proposed in
Refs.~\cite{SMP,SMPC} (if we ignore the influence of optical losses and
other noise sources). 

\subsection{Frequency dependence of coupling constants $\mathcal{K}$, and
Sagnac interferometers as speed meters}

\begin{figure}
\begin{center}

\vspace{0.5cm}
\includegraphics[width=0.45\textwidth]{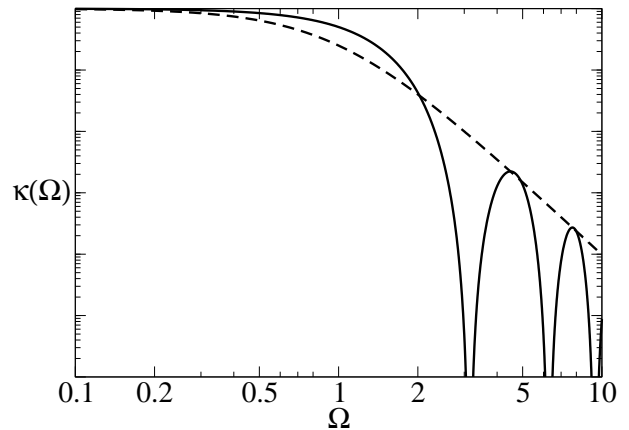}

\caption{The coupling constant \protect$ \mathcal{K}_{\rm sagnac}(\Omega )\protect$
for non-signal-recycled DL [in solid line, Eq.~(\ref{eq:KSRapp}) with
$\rho=0$, $\tau=1$] and FP [in dashed line, Eq.~(\ref{eq:KFPsagnacSR})
    with $\delta=2\Omega_{\rm s}$] sagnac
interferometers, in arbitrary (logarithmic) units, with \protect$
\Omega \protect$ 
measured in units of \protect$ \gamma _{\mathrm{DL}}\protect$
(DL case) and \protect$ \Omega _{\mathrm{s}}\protect$ (FP case),
respectively. \label{Fig:KnonSR}}
\end{center}
\end{figure}

\begin{figure*}
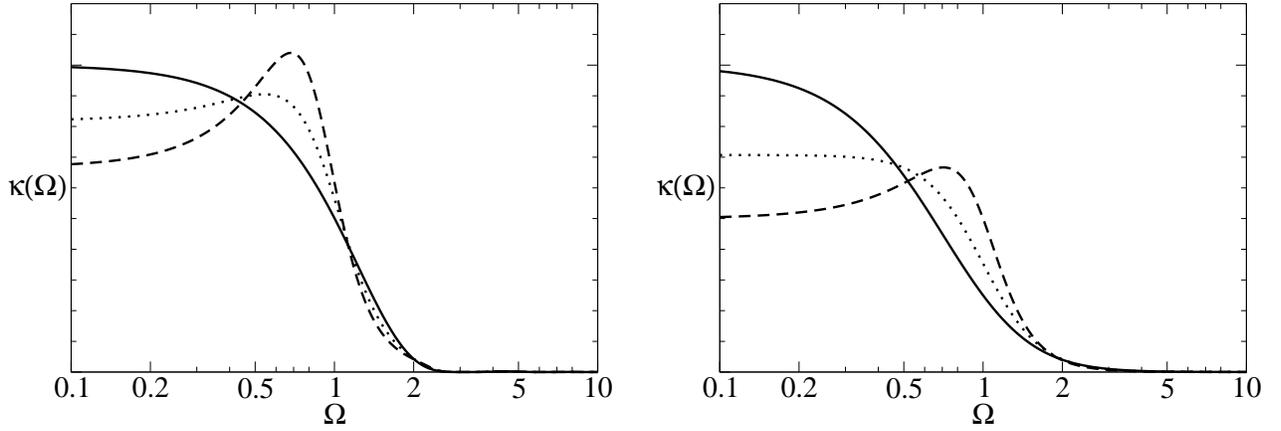

\begin{center}
\begin{tabular}{cc}
\includegraphics[width=0.45\textwidth]{Fig4a.eps} \hspace{0.25cm}& \hspace{0.2cm}
\includegraphics[width=0.45\textwidth]{Fig4b.eps}
\end{tabular}
\end{center}

\caption{The coupling constant \protect$ \mathcal{K}_{\rm sagnac\,SR}(\Omega )\protect$
for signal-recycled DL [left panel, Eq.~(\ref{eq:KSRapp})] and FP [right
  panel, Eq.~(\ref{eq:KFPsagnacSR})] Sagnac interferometers,
in arbitrary (linear) units, with \protect$ \Omega \protect$ measured in
units of \protect$ \gamma _{\mathrm{DL}}\protect$ (DL case) and
\protect$ \Omega _{\mathrm{s}}\protect$ (FP case), respectively.
For DL: cases with \protect$ \rho =0\protect$ (solid curve), 0.1
(dotted curve) and 0.2 (dashed curve) are
plotted . For FP cases with \protect$ \delta =2\Omega
_{\mathrm{s}}\protect$ (solid curve),
\protect$ \sqrt{2}\Omega _{\mathrm{s}}\protect$ (dotted curve), and
\protect$ \Omega _{\mathrm{s}}\protect$ (dashed curve) 
are plotted, corresponding to \protect$ \rho =0\protect$, 0.172,
and 0.333. 
\label{Fig:KSR}}
\end{figure*}

As can be seen both analytically in Eqs. (\ref{eq:KFPsagnacSR})
and (\ref{eq:delta}) and graphically in Fig.~\ref{Fig:KnonSR},
the coupling constant $ \mathcal{K}_{\rm sagnac}$  of a Sagnac interferometer
without signal recycling (i.e., $ \tau =1$, $ \rho =0$) approaches
a constant
as $ \Omega \rightarrow 0$,
which also turns out to be its maximum. 
This fact, combined with the input-output relation (\ref{eq:inoutsagnacgeneric2}),
suggests that the second output quadrature $ q_{2}$ is indeed
sensitive to the speed of the interferometer induced by the gravitational
wave, since at low frequencies 
\begin{equation}
q_{2}(\mathrm{signal}\, \mathrm{part})\propto \Omega
\sqrt{\mathcal{K}(\Omega =0)}
\tilde{x}^{\mathrm{GW}}\propto
\mathrm{momentum}\, .
\end{equation} 
(A more detailed discussion of the link between $ \mathcal{K}$
and a speed meter's performance is given in Sec.~IIIA of Ref.~\cite{SMPC};
that discussion, in the framework of a Michelson speed meter, is equally
valid for a Sagnac speed meter.) When signal recycling is added, the shape of $ \mathcal{K}(\Omega )$
can be adjusted for optimization purposes; examples are shown
in Fig.~\ref{Fig:KSR}.

\section{Noise spectral density\label{sec:spectraldensity}}

In this section, we shall assume that homodyne detection can be carried
out on any (frequency-independent) quadrature, 
\begin{equation}
\label{eq:qPhi}
q_{\Phi }=q_{1}\cos \Phi +q_{2}\sin \Phi \, .
\end{equation}
Homodyne detection is essential for QND interferometers, if they are 
to beat the SQL by substantial amounts; the additional
noise associated with heterodyne detection schemes can seriously limit
an interferometer's ability to beat the SQL \cite{rfmod}. 

The noise
spectral density associated with the input-output relations (\ref{eq:inoutsagnacgeneric1})
and (\ref{eq:inoutsagnacgeneric2}) can be obtained in a manner analogous
to that of Sec.~IV of KLMTV or Sec.~III of Ref.~\cite{SMPC}.
The result is
\begin{equation}
\label{eq:ShPhi}
S_{h}=\left[ \frac{(\cot \Phi -\mathcal{K}_{\mathrm{sagnac}\,
      \mathrm{SR}})^{2}+1}{2\mathcal{K}_{\mathrm{sagnac}\,
      \mathrm{SR}}}\right] h_{\mathrm{SQL}}^{2}\, . 
\end{equation}
As is also discussed in Refs.~\cite{SMBGKT,SMP,SMPC},
the optimal quadrature to observe is the one with 
\begin{equation}
\label{eq:cotPhi}
\cot \Phi =\mathcal{K}_{\mathrm{max}}\equiv \max _{\Omega
}\mathcal{K}_{\mathrm{sagnac}\, \mathrm{SR}}(\Omega )\, ; 
\end{equation}
for this quadrature the noise spectral density is 
\begin{equation}
\label{eq:ShOpt}
S_{h}=\left[
  \frac{(\mathcal{K}_{\mathrm{max}}-\mathcal{K}_{\mathrm{sagnac}\,
      \mathrm{SR}})^{2}+1}{2\mathcal{K}_{\mathrm{sagnac}\,
      \mathrm{SR}}}\right] h_{\mathrm{SQL}}^{2}\, . 
\end{equation}

\begin{figure*}
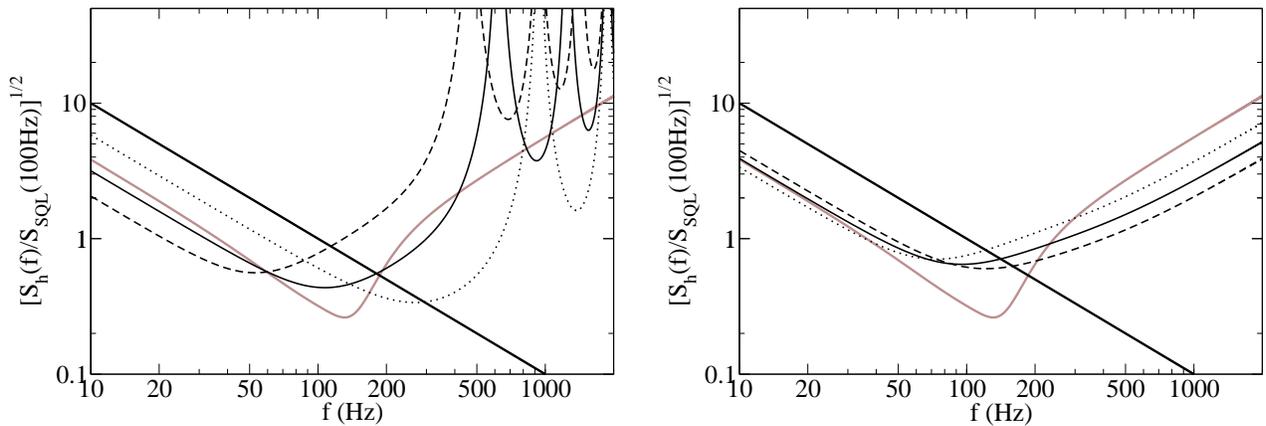

\begin{center}
\begin{tabular}{cc}
\includegraphics[width=0.45\textwidth]{Fig5a.eps} \hspace{0.25cm}& \hspace{0.2cm}
\includegraphics[width=0.45\textwidth]{Fig5b.eps}
\end{tabular}
\end{center}

\caption{The noise spectral density [for the optimized quadrature,
  see Eq.~(\ref{eq:cotPhi})] of non-signal-recycled DL (left panel) and
FP (right panel) Sagnac interferometers [Eq.~(\ref{eq:ShOpt}), setting
  $\rho=1$ and $\tau=0$], assuming  \protect$ I_{\mathrm{c}}=8.2\, \mathrm{MW}\protect$
and \protect$ m=40\, \mathrm{kg}\protect$. 
{[}By injecting
squeezed vacuum (with squeeze factor $e^{-2R}$) into the dark port, one 
can reduce \protect$ I_{\mathrm{c}}\protect$
by a factor \protect$ e^{2R}\sim 10\protect$.{]} 
For DL: cases
with \protect$ \mathcal{B}=40\protect$ (dotted curve), 60
(solid curve) and 80 (dashed curve) are plotted. For FP: cases
with \protect$ \Omega _{\mathrm{s}}=2\pi \times 200\, \mathrm{Hz}\protect$
(dotted curve), \protect$ 2\pi \times 220\, \mathrm{Hz}\protect$
(solid curve) and \protect$ 2\pi \times 240\, \mathrm{Hz}\protect$
(dashed curve) are plotted. The noise curves for the fiducial speed
meter (in gray) and the SQL (dark straight lines) are also plotted
in both panels for comparison. \label{Fig:ShNonSR}}
\end{figure*}

In the left panel of Fig.~\ref{Fig:ShNonSR}, we plot the noise spectral
density of a delay-line Sagnac interferometer  {\em without} signal recycling,
with $ m=40\, \mathrm{kg}$ and $ I_{\mathrm{c}}=8.2\, \mathrm{MW}$
(the characteristic circulating power used for the Michelson speed meter 
in Refs.~\cite{SMP,SMPC}),
and with $ \mathcal{B}=$40, 60, 80 [corresponding to powers in a single
beam equal to $ 102.5\, \mathrm{kW}$, $ 68.3\, \mathrm{kW}$
and $ 51.3\, \mathrm{kW}$, respectively; see Eq.~(\ref{eq:Ib});
these powers can be lowered by injecting squeezed vacuum into
the dark port, as we shall discuss below]. The noise spectral density
of the fiducial Michelson speed meter of Refs.~\cite{SMP,SMPC}, with
the same $ I_{\mathrm{c}}$ and $ m$, and (in their notation)
$ \Omega =2\pi \times 173\, \mathrm{Hz}$, $ \delta =2\pi \times 200\, \mathrm{Hz}$,
is also plotted for comparison. In the right panel, we plot the noise
spectral density of a ring-cavity Sagnac interferometer {\em without} signal
recycling, with the same $ m$ and $ I_{\mathrm{c}}$, and with
$ \gamma _{\mathrm{FP}}=2\pi \times 200\, \mathrm{Hz}$, $ 2\pi \times 220\, \mathrm{Hz}$
and $ 2\pi \times 240\, \mathrm{Hz}$. As one can see in the two
panels, both configurations of non-recycled Sagnac interferometers
exhibit broadband QND performance, with the beating of the SQL concentrated
at low frequencies.

\begin{figure*}
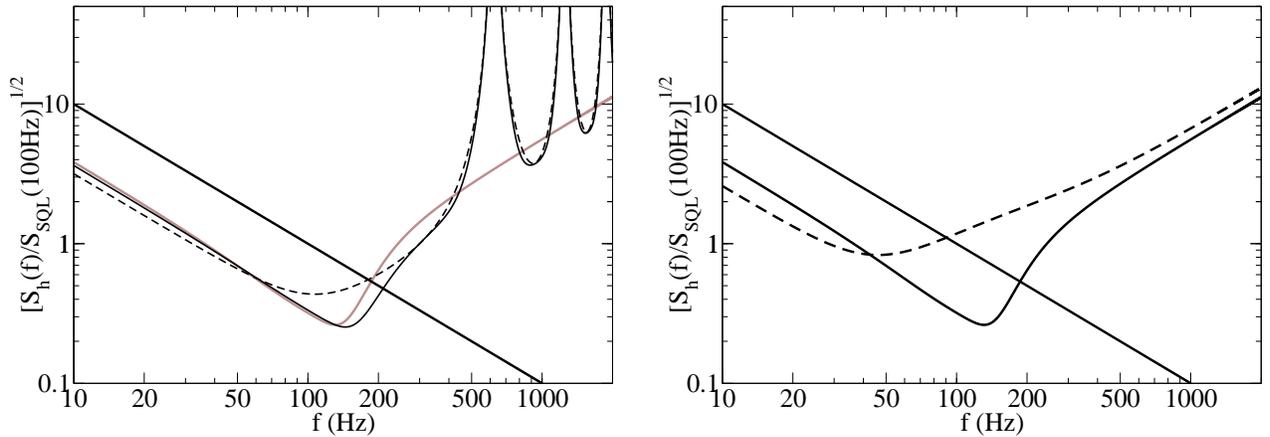

\begin{center}
\begin{tabular}{cc}
\includegraphics[width=0.45\textwidth]{Fig6a.eps} \hspace{0.25cm}& \hspace{0.2cm}
\includegraphics[width=0.45\textwidth]{Fig6b.eps}
\end{tabular}
\end{center}

\caption{The solid curves are the noise spectral densities of 
signal-recycled DL (left panel) and FP
(right panel) Sagnac interferometers [Eq.~(\ref{eq:ShOpt})], assuming \protect$ I_{\mathrm{c}}=8.2\, \mathrm{MW}\protect$
and \protect$ m=40\, \mathrm{kg}\protect$. 
{[}By injecting squeezed vacuum into the dark port,
one can reduce \protect$ I_{\mathrm{c}}\protect$ by a factor \protect$ e^{2R}\sim 10\protect$.{]}
For DL, we take \protect$ \mathcal{B}=60\protect$, \protect$ \rho =0.09\protect$;
for FP, we take \protect$ \Omega _{\mathrm{s}}=2\pi \times 173\, \mathrm{Hz}\protect$
and \protect$ \delta =2\pi \times 200\, \mathrm{Hz}\protect$,
which correspond to \protect$ T=0.0564\protect$ and \protect$ \rho =0.268\protect$.
The corresponding non-recycled noise curves are also shown, as dashed
curves. The noise curve of the fiducial 
Michelson speed meter is plotted in gray in the left panel and is identical to the 
solid, signal-recycled FP Sagnac noise curve in the right panel;
the standard quantum limit is shown as dark straight lines.\label{Fig:ShSR}}
\end{figure*}

Signal recycling allows us to improve and optimize the Sagnac interferometers
so they have similar performance to a Michelson speed meter; i.e., so
they beat
the SQL by a roughly constant factor over a substantially broader 
frequency band than without signal recycling.  
In particular, since the spectral density (\ref{eq:ShPhi}) only depends
on $ \mathcal{K}$, and $ \mathcal{K}_{\mathrm{sagnac}\, \mathrm{SR}}^{\mathrm{FP}}$
is the same as that of a Michelson speed meter, the signal-recycled Sagnac
interferometers with ring cavities will have the same performance
as the Michelson speed meters. In Fig.~\ref{Fig:ShSR}, we give one
example for each of the DL and FP configurations. In the left panel
we plot the noise spectral density for a signal-recycled DL Sagnac
with $ m=40\, \mathrm{kg}$, $ I_{\mathrm{c}}=8.2\, \mathrm{MW}$,
$ \mathcal{B}=60$ (and therefore $ I_{\mathrm{b}}=68\, \mathrm{kW}$)
and $ \rho =0.09$ (dark solid curve), compared with that of the
corresponding non-recycled ($ \rho =0$) interferometer (dashed
curve), and that of the fiducial Michelson speed meter (gray solid
curve). In the right panel we plot the the noise spectral density
of a signal-recycled FP Sagnac interferometer with $ T=0.0564$,
$ \rho =0.268$, corresponding to $ \Omega _{\mathrm{s}}=2\pi \times 173\, \mathrm{Hz}$,
and $ \delta =2\pi \times 200\, \mathrm{Hz}$ {[}from Eq.~(\ref{eq:delta}){]}.
This interferometer has the same noise spectral density as the fiducial
Michelson speed meter. {[}The two noise curves agree perfectly, appearing
as the solid curve in the panel.{]} The corresponding non-recycled
noise curve (with $ \rho =0$) is also plotted (the dashed curve)
for comparison.

As conceived by Caves \cite{Caves}  and discussed in
Refs.~\cite{KLMTV,SMPC}, injecting squeezed vacuum into the dark-port
can lower the required circulating power. For example, as discussed
in Sec.~IVA of Ref.~\cite{SMPC}, for speed meters with input-output
relations with the form of Eqs.~(\ref{eq:inoutsagnacgeneric1}) and
(\ref{eq:inoutsagnacgeneric2}), the circulating power can be lowered
by the squeeze factor $ e^{-2R}$, while maintaining the same
performance. In the LIGO-III era, it is reasonable to
expect $ e^{-2R}\sim 0.1$ \cite{KLMTV},
so the circulating powers cited above can be lowered by a factor $ \sim 10$. 
The resulting fiducial circulating power, $I_c = 8.2 \mathrm{MW}/10 = 
820\mathrm{kW}$ is about the same as planned for the second-generation
Advanced LIGO interferometers.

Finally, for signal-recycled FP Sagnac interferometers, since they are 
equivalent to the Michelson speed meters of Ref.\ \cite{SMPC}, 
one can further improve the high-frequency performance by performing
frequency-dependent homodyne detection with the aid of two
kilometer-scale optical filters at the dark-port output; 
see Sec.~IVB of \cite{SMPC}.

\section{Discussion of technical issues}

\label{sec:furtherissues} 

We shall now comment on three technical issues that might affect
significantly the performances of Sagnac speed-meter interferometers:

{\em Optical Losses.}
So far in this paper, we have regarded all interferometers as ideal; most
importantly, we have ignored optical losses. As has been shown by
several studies of the Michelson case \cite{KLMTV,SMPC}, optical
losses can sometimes be the limiting factor on the sensitivity of 
a QND interferometer.
However, as shown in Ref.~\cite{SMPC}, Michelson speed meters are
\emph{less susceptible to} optical losses than Michelson position
meters (even though the losses may be enhanced by the larger number of
optical surfaces on which to scatter or absorb, and by the fact that the 
coupling constant $ \mathcal{K}(\Omega )$
remains finite as $ \Omega \rightarrow 0$ rather than growing
to infinity). It is plausible that this feature will be retained, at
least for optical losses associated with the individual optical
elements, and with the readout scheme, but
rigorous calculations are yet to be carried out.  Moreover, the losses
due to the use of  diffractive optics and polarization techniques
in some Sagnac configurations \cite{AllReflective} deserve serious
study.

{\em High Power Through the Beam Splitter.}
As we saw at the end of Sec.~\ref{sec:spectraldensity},
for FP Sagnac interferometers, in order to optimize the shape of the noise
curve, the required values of the power transmissivity of the ITM
can become as large as $ 0.05$, which may require optical powers
at the level of tens of kilowatts through the beamsplitter (even when
squeezed vacuum is injected into the dark port); this may pose
a problem for implementation. In Michelson speed meters,
a resonant-side-band-extraction technique can be used to greatly reduce the power through
the beamsplitter without affecting the interferometer's performance, but 
it is not
clear whether an analogous trick exists for Sagnac interferometers. 

{\em Susceptibility to Mirror Tilt and Imperfections.}
In the low-laser-power limit, the Sagnac interferometer
is known to be more susceptible to mirror tilting than are Michelson 
interferometers,
but less susceptible to geometric imperfections of mirrors \cite{PGM98}.
A study of these susceptibilities needs to be carried out in the context of 
high laser power,
in order to see whether they pose any serious difficulty in the
implementation of Sagnac speed meters.

\section{Conclusions}

In this paper, a quantum-mechanical study of idealized Sagnac interferometers,
including radiation-pressure effects, has been carried out. As was already 
known, Sagnac interferometers are sensitive only to the time varying part
of the antisymmetric mode of mirror displacement. It was a short and trivial
step, in this paper, to demonstrate that this means a Sagnac interferometer
measures the test masses' relative speed or momentum and therefore
is a speed meter with QND capabilities.  Detailed computations revealed 
that, as for other speed meters, a broad-band QND performance can be obtained, 
when frequency-independent
homodyne detection is performed at the dark port. Signal recycling
can be employed to further optimize the noise spectrum so it is comparable
to that of a Michelson speed meter (or exactly the same, for 
Sagnac configurations with ring cavities in the arms); and, by contrast
with the Michelson, this can be achieved without the need
for any additional kilometer-scale FP cavity. {[}In the case of frequency-dependent
homodyne detection with the aid of two kilometer-scale filter cavities, the 
Sagnac speed meter still needs one less optical cavity
than its Michelson counterpart.{]} If further technical
issues, including those related to optical losses (Sec.~\ref{sec:furtherissues}),
can be resolved, the Sagnac optical topology will be a strong
candidate for third-generation gravitational-wave interferometers,
such as LIGO-III and EURO. 

\begin{acknowledgments}
I thank Kip Thorne for discussions, comments and suggestions on the
manuscript, and encouragement to finish it. This research
was supported in part by NSF grant PHY-0099568, and by the Barbara and David
Groce Fund of the San Diego Foundation. 
\end{acknowledgments}
\appendix

\section{Input-output relations for the arms}

Since the north and east arms are identical, we need only analyze
one of them. For concreteness, we study the East arm. In
Appendix~B of Ref.~\cite{KLMTV}, KLMTV derived the input-output
relation for a simple FP cavity, using the same Caves-Schumaker quadrature
formalism \cite{Caves_Schumaker} as we use here. The input-output relations 
for optical delay-line arms and ring cavities can be derived analogously: 

\subsection{Optical delay line}

Following the procedure of KLMTV, we initially suppose that the ITM is not
moving, and we denote the displacement of the ETM by $ x_{\mathrm{E}}(t)$.
Suppose the R beam has an electric field amplitude
\begin{widetext} 
\begin{equation}
\label{eq:DLinput}
E^{\mathrm{RE}\, \mathrm{in}}(t)=[D+E^{\mathrm{RE}\, \mathrm{in}}_{1}(t)]\cos \omega _{0}t+E^{\mathrm{RE}\, \mathrm{in}}_{2}(t)\sin \omega _{0}t
\end{equation}
at the location where it enters the E arm; here 
$D$ is the (classical) carrier amplitude and 
$ E^{\mathrm{RE}\, \mathrm{in}}_{1,2}(t)$
are the sideband quadrature fields
\begin{equation}
\label{eq:EREinquad}
E^{\mathrm{RE}\, \mathrm{in}}_{1,2}(t)=\sqrt{\frac{4\pi \hbar \omega
    _{0}}{{\mathcal{A}}c}}\int _{0}^{+\infty }\frac{d\Omega }{2\pi
}\left[ a^{\mathrm{RE}}_{1,2}e^{-i\, \Omega \, t}+h.c.\right] \, , 
\end{equation}
with ``h.c.'' meaning ``Hermitian conjugate''.
The output beam after $ \mathcal{B}$ bounces is delayed by 
\begin{equation}
\Delta t=2\mathcal{B}L/c+2\big[
  x_{\mathrm{E}}(t-L/c)+x_{\mathrm{E}}(t-3L/c)+\cdots
  +x_{\mathrm{E}}(t-(2\mathcal{B}-1)L/c)\big]\,,
\end{equation} 
so 
\begin{eqnarray}
E^{\mathrm{RE}\, \mathrm{out}}(t)=E^{\mathrm{RE}\,
 \mathrm{in}}(t-\Delta t) & \approx  & [D+E^{\mathrm{RE}\,
 \mathrm{in}}_{1}(t-2\mathcal{B}L/c)]\cos \omega
 _{0}t+E^{\mathrm{RE}\, \mathrm{in}}_{2}(t-2\mathcal{B}L/c)\sin \omega
 _{0}t\nonumber \\ 
 & + & \frac{2\omega _{0}D}{c}\sum 
 _{k=1}^{\mathcal{B}}x_{\mathrm{E}}\left( t-(2k-1)L/c\right) \,
 .\label{eq:EREout}  
\end{eqnarray} 
Comparing with 
\begin{equation}
\label{eq:EREoutquad}
E^{\mathrm{RE}\, \mathrm{out}}_{1,2}(t)=\sqrt{\frac{4\pi \hbar \omega
    _{0}}{{\mathcal{A}}c}}\int _{0}^{+\infty }\frac{d\Omega }{2\pi
}\left[ b^{\mathrm{RE}}_{1,2}e^{-i\, \Omega \, t}+h.c.\right] \, , 
\end{equation}
we obtain
\begin{eqnarray}
b^{\mathrm{RE}}_{1} & = & e^{2i\mathcal{B}\Omega L/c}a_{1}^{\mathrm{RE}}\, ,\label{eq:DLinout1} \\
b_{2}^{\mathrm{RE}} & = & e^{2i\mathcal{B}\Omega
  L/c}a_{2}^{\mathrm{RE}}
+\frac{2\omega
  _{0}}{c}\sqrt{\frac{2I_{\mathrm{b}}}{\hbar \omega _{0}}}\left(
\frac{\sin \mathcal{B}\Omega L/c}{\sin \Omega L/c}\right)
e^{i\mathcal{B}\Omega L/c}\tilde{x}_{\mathrm{E}}\,
.\label{eq:DLinout2}  
\end{eqnarray}
where $\tilde{x}_{\mathrm{E}}$ is the Fourier transform of
$x_{\mathrm{E}}(t)$. 
Here $ I_{\mathrm{b}}$ is the power of the beam, 
\begin{equation}
\label{eq:Ic}
I_{\mathrm{b}}=\frac{D^{2}{\mathcal{A}}c}{8\pi }\, ,
\end{equation}
 which is related to the total circulating power by 
\begin{equation}
\label{eq:Ib}
I_{\mathrm{b}}=\frac{I_{\mathrm{c}}}{2\mathcal{B}}\, .
\end{equation}
The physical meanings of Eqs.~(\ref{eq:DLinout1}) and (\ref{eq:DLinout2})
can be roughly explained as follows: 
(i) the gravitational-wave signal embodied
in $ \tilde{x}_{\mathrm{E}}$ is only present in the second (phase)
quadrature, $ b^{\mathrm{RE}}_{2}$, of the output sideband field,
i.e., in the second term of the right-hand side of Eq.~(\ref{eq:DLinout2});
(ii) the first term on the right-hand sides of Eqs.~(\ref{eq:DLinout1})
and (\ref{eq:DLinout2}) represents the shot noise, which originates
from the quantum fluctuations of the input field. Obviously, the relations
(\ref{eq:DLinout1}) and (\ref{eq:DLinout2}) also apply to the L
beam, with the change of superscript R to L. 

Next we must study the motion of the end mirror, which is influenced
by both the passing gravitational wave and the radiation-pressure
force:
\begin{equation}
\label{eq:xE}
x_{\mathrm{E}}=x^{\mathrm{GW}}_{\mathrm{E}}+x_{\mathrm{E}}^{\mathrm{BA}}\,
,\quad \ddot{x}_{\mathrm{BA}}=\frac{1}{m}F_{\mathrm{RP}}. 
\end{equation}
Here $ x_{\mathrm{BA}}$ is the displacement induced by the radiation-pressure
force, or the \emph{back action} of the measurement process, which
eventually gives rise to the \emph{radiation-pressure noise.} The
radiation-pressure force $ F_{\mathrm{RP}}$ comes from both the
L and R beams: 
\begin{equation}
\label{eq:FRPexact}
F_{\mathrm{RP}}(t)=\frac{{\mathcal{A}}}{2\pi }\sum
_{k=1}^{\mathcal{B}}\left\{ \left[ E^{\mathrm{RE}\,
    \mathrm{in}}(t-(2k-1)L/c)\right] ^{2}+\left[ E^{\mathrm{LE}\,
    \mathrm{in}}(t-(2k-1)L/c)\right] ^{2}\right\} \, . 
\end{equation}
 However, we are only interested in the \emph{fluctuating} and 
\emph{low-frequency} part (in the gravitational-wave band) of the force, which
comes from the beating of the sideband fields against the carrier:
\begin{equation}
\label{eq:FRPfluc}
F^{\mathrm{fluc}}_{\mathrm{RP}}(t)=\frac{D{\mathcal{A}}}{2\pi }\sum
_{k=1}^{\mathcal{B}}\left[ E_{1}^{\mathrm{RE}\,
    \mathrm{in}}(t-(2k-1)L/c)+E_{1}^{\mathrm{LE}\,
    \mathrm{in}}(t-(2k-1)L/c)\right] \, . 
\end{equation}
Combining Eqs.~(\ref{eq:xE}) and (\ref{eq:FRPfluc}) and transforming
into the frequency domain, we obtain the Fourier transform of the mirror
displacement in the GW frequency band {[}note that Eq.~(\ref{eq:Ic})
is used again{]}: 
\begin{equation}
\label{eq:xERP}
\tilde{x}_{\mathrm{E}}=\tilde{x}_{\mathrm{E}}^{\mathrm{GW}}-\frac{4}{m\Omega
  ^{2}c}\sqrt{2\hbar \omega _{0}I_{\mathrm{b}}}\left( \frac{\sin
  \mathcal{B}\Omega L/c}{\sin \Omega L/c}\right) e^{i\mathcal{B}\Omega
  L/c}\left( \frac{a^{\mathrm{RE}}_{1}+a^{\mathrm{LE}}_{1}}{2}\right)
\, . 
\end{equation}
This, when combined with Eq.~(\ref{eq:DLinout2}), yields:
\begin{equation}
\label{eq:b2RERP}
b_{2}^{\mathrm{RE}}=e^{2iB\Omega
  L/c}a_{2}^{\mathrm{RE}}-\frac{16I_{\mathrm{b}}\omega _{0}}{m\Omega
  ^{2}c^{2}}\left( \frac{\sin \mathcal{B}\Omega L/c}{\sin \Omega
  L/c}\right) ^{2}e^{2i\mathcal{B}\Omega L/c}\left(
\frac{a_{1}^{\mathrm{RE}}+a_{1}^{\mathrm{LE}}}{2}\right)
+\frac{2\omega _{0}}{c}\sqrt{\frac{2I_{\mathrm{b}}}{\hbar \omega
    _{0}}}\left( \frac{\sin \mathcal{B}\Omega L/c}{\sin \Omega
  L/c}\right) \tilde{x}_{\mathrm{E}}^{\mathrm{GW}}\, . 
\end{equation}
The second term on the right-hand side is the radiation-pressure noise.
In reality, the internal mirrors (ITM's), the beamsplitter, and the connection
mirror will also move under the radiation-pressure force (but they
are \emph{not} influenced by gravitational waves). When $ \mathcal{B}\gg 1$
and $ \Omega L/c\ll 1$, only the internal mirrors need be taken
into account, and the effect is just a doubling of the radiation pressure
noise in Eq.~(\ref{eq:b2RERP}). Hence we arrive at the complete input-output
relation for the East arm, put into a more compact form 
(similar to those in KLMTV
\cite{KLMTV}):
\begin{eqnarray}
b^{\mathrm{RE}}_{1} & = & e^{2i\Psi _{\mathrm{DL}}}a^{\mathrm{RE}}_{1}\, ,\label{eq:REinoutDL1} \\
b_{2}^{\mathrm{RE}} & = & e^{2i\Psi _{\mathrm{DL}}}\left[
  a_{2}^{\mathrm{RE}}-\mathcal{K}_{\mathrm{DL}}(a_{1}^{\mathrm{RE}}+a^{\mathrm{LE}}_{1})\right]+
e^{i\Psi
  _{\mathrm{DL}}}\frac{\sqrt{2\mathcal{K}_{\mathrm{DL}}}}{h_{\mathrm{SQL}}}\sqrt{2}\, \tilde{x}^{\mathrm{GW}}_{E}/L\, ;\label{eq:REinoutDL2}      
\end{eqnarray} 
where
\begin{eqnarray}
\Psi _{\mathrm{DL}} & = & \mathcal{B}\Omega L/c\, ,\label{eq:PsiDLapp} \\
\mathcal{K}_{\mathrm{DL}} & = & \frac{16I_{\mathrm{b}}\omega
  _{0}}{m\Omega ^{2}c^{2}}\left( \frac{\sin \mathcal{B}\Omega
  L/c}{\sin \Omega L/c}\right) ^{2} =\frac{8I_{\mathrm{c}}\omega
  _{0}}{m\mathcal{B}\Omega ^{2}c^{2}}\left( \frac{\sin
  \mathcal{B}\Omega L/c}{\sin \Omega L/c}\right) ^{2}\,
.\label{eqKDLapp}  
\end{eqnarray}
 The input-output relation for the L beam can be obtained by exchanging
RE and LE in Eqs.~(\ref{eq:REinoutDL1}) and (\ref{eq:REinoutDL2}).

\subsection{Ring cavity}

Again, let us consider the East arm. Suppose again, for the moment,  
that only the ETM is allowed to move. Then the input-output relations
for the fields immediately inside the ITM can be obtained easily from
the results for optical delay lines {[}Eqs.~(\ref{eq:REinoutDL1})--(\ref{eqKDLapp}),
with a factor $ 1/2$ multiplying the radiation-pressure noise term,
since again as a first step we are only allowing the ETM to move{]}:
\begin{eqnarray}
B_{1}^{\mathrm{RE}} & = & e^{2i\Omega L/c}A_{1}^{\mathrm{RE}}\, ,\label{eq:onebounce1} \\
B_{2}^{\mathrm{RE}} & = & e^{2i\Omega L/c}\left[
  A_{2}^{\mathrm{RE}}-\mathcal{K}^{\mathcal{B}=1}_{\mathrm{DL}}\left(
  \frac{A_{1}^{\mathrm{RE}}+A_{1}^{\mathrm{LE}}}{2}\right) \right]
+e^{i\Omega
  L/c}\frac{\sqrt{2\mathcal{K}_{\mathrm{DL}}^{\mathcal{B}=1}}}{h_{\mathrm{SQL}}}\sqrt{2}\, \tilde{x}_{\mathrm{E}}/L\, ;  
\end{eqnarray} 
where 
\begin{equation}
\label{eq:Konebounce}
\mathcal{K}_{\mathrm{DL}}^{B=1}=\frac{8I_{\mathrm{c}}\omega _{0}}{m\Omega ^{2}c^{2}}\, .
\end{equation}
As before, the input-output relation for the L beam is obtained by
exchanging RE and LE. The fields outside the ITM are related to these
fields by
\begin{equation}
\label{eq:FPR}
b^{\mathrm{RE}}_{1,2}=-\sqrt{R}a_{1,2}^{\mathrm{RE}}+\sqrt{T}B^{\mathrm{RE}}_{1,2}\,
,\quad \quad
A^{\mathrm{RE}}_{1,2}=\sqrt{T}a_{1,2}^{\mathrm{RE}}+\sqrt{R}B^{\mathrm{RE}}_{1,2}\,
; 
\end{equation}

\begin{equation}
\label{eq:FPL}
b^{\mathrm{LE}}_{1,2}=-\sqrt{R}a_{1,2}^{\mathrm{LE}}+\sqrt{T}B^{\mathrm{LE}}_{1,2}\,
,\quad \quad
A^{\mathrm{LE}}_{1,2}=\sqrt{T}a_{1,2}^{\mathrm{LE}}+\sqrt{R}B^{\mathrm{LE}}_{1,2}\,
. 
\end{equation}
Combining Eqs.~(\ref{eq:onebounce1})--(\ref{eq:FPL}), we obtain\begin{eqnarray}
b_{1}^{\mathrm{RE}} &  & =\frac{e^{2i\Omega L/c}-\sqrt{R}}{1-e^{2i\Omega L/c}\sqrt{R}}a_{1}^{\mathrm{RE}}\, ,\label{eq:inoutFP1} \\
b_{2}^{\mathrm{RE}} &  & =\frac{e^{2i\Omega
    L/c}-\sqrt{R}}{1-e^{2i\Omega L/c}\sqrt{R}}\left[
  a_{2}^{\mathrm{RE}}
-\frac{T\mathcal{K}_{\mathrm{DL}}^{\mathcal{B}=1}}{1-2\sqrt{R}\cos
  (2\Omega L/c)+R}\left(
\frac{a_{1}^{\mathrm{RE}}+a_{1}^{\mathrm{LE}}}{2}\right) \right]
\nonumber \\    
 &  & +\frac{e^{i\Omega L/c}\sqrt{T}}{1-e^{2i\Omega
    L/c}\sqrt{R}}\frac{\sqrt{2\mathcal{K}_{\mathrm{DL}}^{\mathcal{B}=1}}}{h_{\mathrm{SQL}}}\sqrt{2}\, \tilde{x}_{\mathrm{E}}/L\, .\label{eq:inoutFP2}  
\end{eqnarray}
As before, the first terms on the right-hand sides of Eqs.~(\ref{eq:REinoutFP1})
and (\ref{eq:REinoutFP2}) represent the shot noise, the second term
on the right-hand side of Eq.~(\ref{eq:REinoutFP2}) represents the
radiation-pressure noise, and the third term on the right-hand side
of Eq.~(\ref{eq:REinoutFP2}) is the gravitational-wave signal. Again,
other optical elements besides the ETM can also be influenced by the
radiation-pressure force; but when $ T\ll 1$, we need only consider
the radiation-pressure force on the ITM and the other cavity mirror
near the ITM. Suppose all three sides of the ring cavity are on
resonance with the carrier frequency.  Then it is obvious that, at
leading order in $ \Omega L/c$ and $ T$, the fluctuating forces
associated with the beams at the locations of the three mirrors are
the same. However, since the in-cavity light is incident on the two
near mirrors at $ 45^{\circ }$, the motions of each of them induced
by the radiation pressure is $ 1/\sqrt{2}$ that of the ITM, in
the direction normal to their surfaces. Also because their motion
directions are again $ 45^{\circ }$ to the propagation direction
of the beams, the resulting radiation-pressure noise is reduced by
an additional $ 1/\sqrt{2}$. In the end, the net radiation-pressure
noise due to the two near mirrors is equal to that due to the end
mirror. Doubling the radiation-pressure noise in Eq.~(\ref{eq:REinoutFP2}),
we obtain the input-output relation of the ring cavity, which we put into a form
similar to that of the optical delay line:
\begin{eqnarray}
b^{\mathrm{RE}}_{1} & = & e^{2i\Psi _{\mathrm{FP}}}a^{\mathrm{RE}}_{1}\, ,\label{eq:REinoutFP1} \\
b_{2}^{\mathrm{RE}} & = & e^{2i\Psi _{\mathrm{FP}}}\left[
  a_{2}^{\mathrm{RE}}
-\mathcal{K}_{\mathrm{FP}}(a_{1}^{\mathrm{RE}}+a^{\mathrm{LE}}_{1})\right]
+e^{i\Psi
  _{\mathrm{FP}}}\frac{\sqrt{2\mathcal{K}_{\mathrm{FP}}}}{h_{\mathrm{SQL}}}\sqrt{2}\, \tilde{x}^{\mathrm{GW}}_{E}/L\, ;\label{eq:REinoutFP2}     
\end{eqnarray} 
with
\begin{eqnarray}
\Psi _{\mathrm{FP}} & = & \arg \frac{e^{i\Omega L/c}}{1-e^{2i\Omega
    L/c}\sqrt{R}} =\arctan \left( \frac{1+\sqrt{R}}{1-\sqrt{R}}\tan
\frac{\Omega L}{c}\right) \, ,\label{eq:PsiKFPapp1} \\ 
\mathcal{K}_{\mathrm{FP}} & = & \left( \frac{T}{1-2\sqrt{R}\cos
    2\Omega L/c+R}\right) \frac{8I_{\mathrm{c}}\omega _{0}}{m\Omega
    ^{2}c^{2}}\, .\label{eq:PsiKFPapp2}  
\end{eqnarray}
\end{widetext}


\begin{thebibliography}{10}
\bibitem{EURO} Although there are not yet detailed papers on EURO or
  LIGO-III, they have been discussed within the gravitational-wave
  community as long-range plans.
\end{thebibliography}
\end{document}